\documentclass[prb,aps,twocolumn,showpacs,preprintnumbers,amsmath,amssymb,floatfix,citeautoscript]{revtex4}

\usepackage{graphicx}
\usepackage{dcolumn}
\usepackage{bm}

\begin{document}


\title{Density functional description of water condensation\\ in proximity of nanoscale asperity}

\author{Pavel B. Paramonov}
 \altaffiliation{Departments of Physics and Polymer Engineering, The University of Akron}
\author{Sergei F. Lyuksyutov}
 \email{sfl@physics.uakron.edu}
 \affiliation{Departments of Physics, Chemistry and Polymer Engineering, The University of Akron, Akron OH 44325}

\date{\today}

\begin{abstract}
We apply non-local density functional formalism to describe an equilibrium distribution of the water-like fluid in the asymmetric nanoscale junction presenting an atomic force microscope (AFM) tip dwelling above an arbitrary surface. The hydrogen bonding dominating in intermolecular attraction is modeled as a square well potential with two adjustable parameters (energy and length) characterizing well's depth and width. A water meniscus formed inside nanoscale junction is explicitly described for different humidity. Furthermore, we suggest a simple approach using polymolecular adsorption isotherms for the evaluation of an energetic parameter characterizing fluid (water) attraction to substrate.  This model can be easily generalized for more complex geometries and effective intermolecular potentials. Our study establishes a framework for the density functional description of fluid with orientational anisotropy induced by the non-uniform external electric field. 
\end{abstract}

\maketitle

\section{Introduction}
\label{sec:intro}

An equilibrium behavior of a fluid near geometrically nonuniform solid surfaces exhibits various peculiarities related to confinement and spatially varying external potentials. A quantitative understanding of water condensation phenomena in proximity of nano-asperities under ambient humidity is important for different areas of research including scanning probe microscopy (SPM), nano-patterning, adhesion and friction at macro- and nanoscale. Specifically, the formation of liquid films and bridges near an asperity of sub-micro- or nano-scale curvature needs to be adequately described. 

An importance of the water condensation phenomenon from ambient atmosphere into nanoscale junction formed by AFM tip dwelling above surface has been recognized in a number of works\cite{Capella_fd_review, Piner_water_eff_imaging, Bloess_nanocell, our_anomalous_current_Si, dip-pen_lithography_general_ref, dip_pen_nanoplotter, our_AFMEN_nature, our_triboelectrification, our_AM_AFMEN, our_zlift_APL}.  The formation of the water meniscus affects strongly the force-distance measurements and AFM imaging resolution\cite{Capella_fd_review, Piner_water_eff_imaging}, provides nanoscale electrochemical cell for SPM oxidation of semiconductors and metals\cite{Bloess_nanocell, our_anomalous_current_Si}, serves as ink transport channel for dip-pen nanolithography\cite{dip-pen_lithography_general_ref, dip_pen_nanoplotter} and plays an essential role in AFM-assisted electrostatic nanolithography (AFMEN) of thin polymeric films\cite{our_AFMEN_nature, our_triboelectrification, our_AM_AFMEN, our_zlift_APL}.   

Direct experimental observations of meniscus formation in the AFM tip - surface junction are complicated due to nanometer-scale size of the region of interest. There are no optical tools available for this purpose due to the wavelength limitation. Indirect estimations of the meniscus's size can be made using SPM oxidation\cite{Bloess_nanocell}, noncontact AFM imaging\cite{Garcia_menics_size_noncont, Garcia_menisc_PRL} and substrate's dissolution in water\cite{menisc_size_from_dissolution}. Several theoretical estimations of the meniscus size and shape based on the macroscopic Kelvin equation\cite{menisc_macro_PRB00, menisc_Gao_APL} and molecular level grand canonical Monte Carlo simulations\cite{menisc_MC_Ratner_JCP02, menisc_MC_Ratner_PRL03, menisc_MC_Ratner_PRL04} have been performed. A macroscopic phenomenological approach based on the modifications of the Kelvin equation has been suggested to describe water attracted to biased AFM tip\cite{Garcia_menisc_PRL}. The model misses molecular level understanding, and is not applicable for the systems confined to the several molecular diameters\cite{capill_condens_MC_Landau, conf_phase_separ_review} typical for AFM tip-substrate separation. Arguably, the approaches based on the Kelvin equation do not take into account strongly adsorbed layers, and neglect the density oscillations near confining surfaces\cite{conf_phase_separ_review}. There is no molecular-level model describing the effects of external electric field to the best of our knowledge.

The goal of this work is to develop a versatile description of the water condensation in proximity of a nanoscale asperity and investigate the influence of spatially non-uniform external electric field, with a focus on AFM nanolithography. Here we concern with the case of isotropic fluid confined in asymmetric nanojunction in the absence of the electric field to develop a basic framework for the case of field-induced orientational anisotropy.     

Typical molecular approaches, applicable to the problem of our interest, include computer simulations of the Monte Carlo\cite{book_MC_Binder} and molecular dynamics\cite{nano_MD_review} types, and a density functional theory (DFT)\cite{DFT_inhomog_fluids_Evans, DFT_in_statmech_Davis}. A grand canonical Monte Carlo simulation of water condensation in the context of the dip-pen nanolithography has been reported in a series of works by Jang {\itshape et.al.}\cite{menisc_MC_Ratner_JCP02, menisc_MC_Ratner_PRL03, menisc_MC_Ratner_PRL04} including clarification of fluctuations influence on the meniscus width. The effects of relative humidity, AFM tip curvative, tip and substrate wetting properties on the meniscus size and capillary force have been studied. 

Unlike other studies, our study concerns with DFT approach to model real systems in which the fluctuations of meniscus's width, as established in\cite{menisc_MC_Ratner_JCP02}, are not significant. A mean field character of the DFT is appropriate since the fluid is considered far from the critical point. Menisci instabilities related to the tip-substrate distance variations dependent on the AFM lever thermal fluctuations, and on other factors could be described in the framework of DFT formulation, however, it will be the subject of a separate study focused on the specifics of AFM-assisted nanolithography.

Another reason for DFT choice with respect to Monte Carlo technique is the convenience of generalization to the case of the electric field induced anisotropy. The Monte Carlo simulations with additional orientational degrees of freedom would require special cautions to avoid local minima and to ensure proper sampling of the complete configurational space, along with increased computational time, especially in the cases of the low acceptance rates for the regions filled with a dense liquid. Molecular dynamics seems to be less convenient as well, because of the variable number of particles dictated by the grand canonical type of the problem, and low densities in the vapor phase potentially providing insufficient number of collisions for proper statistical averaging. Additionally, formulation of the analytical expressions, possible in the framework of DFT, is highly desirable. Finally, another advantage of DFT is a consistency in mapping of the 3D geometry into the quasi-2D description. This avoids errors in quantitative prediction of the solvation forces, present in the case of 2D lattice Monte Carlo simulation \cite{menisc_MC_Ratner_PRL03}.
 
The paper is organized as follows. Section~\ref{sec:model} is entitled ''Model and Methodology''. The first part of the section describes the model including non-local density functional formulation and specifics of fluid, while the second part outlines the computational issues related to the form of the integral equation for density distribution. We discuss multidimensional integrals handling, and describe the iterative procedure developed. The Section~\ref{subsec:bulk} concerns with elucidation of the model parameters of the fluid from the bulk behavior. The Section~\ref{subsec:menisc} presents the results of calculations. The meniscus's size and its variation with respect to the relative humidity and wall separation are discussed. The Section~\ref{subsec:fluid_wall_inter} addresses determination of the fluid-wall interaction parameters to link the model's calculations to particular material systems. The procedure is based on the polymolecular adsorption isotherms. The Section~\ref{sec:summary} presents the summary.

\section{Model and Methodology}
\label{sec:model}

\subsection{Model description}
\label{subsec:model_descr}

\begin{figure}
\includegraphics[width=7cm]{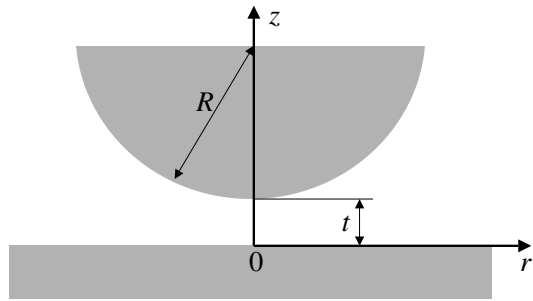}
\caption{Schematic presentation of the nanoscale junction consisting of a planar surface and a spherical asperity with curvature radius $R$, separated by the distance $t$. The fluid is nonuniformly distributed in the junction. The model resembles AFM tip dwelling above a sample surface.}
\label{fig:geom}
\end{figure}

The nano-scale junction consists of planar surface and spherical asperity of the radius $R$, separated from the surface at the distance $t$ (Fig.~\ref{fig:geom}). The choice of the cylindrical coordinate system is related to the azimuthal symmetry. The fluid is nonuniformly distributed in the junction. 

The purpose of the equilibrium DFT applied to the inhomogeneous fluid is to find the spatial distribution for the number density of molecules, $\rho\left(\vec{r}\right)$, by minimizing the grand potential $\Omega$, presented as a functional of $\rho\left(\vec{r}\right)$ for the grand canonical ($T,\;V,\;\mu$) ensemble\cite{DFT_inhomog_fluids_Evans, DFT_in_statmech_Davis}:
\begin{equation}
\label{GCOmega}
	\Omega\left[ \rho\left(\vec{r}\right) \right] = F\left[ \rho\left(\vec{r}\right) \right] + \int \rho\left(\vec{r}\right) \left(V_{ext}\left(\vec{r}\right)-\mu\right) d\vec{r} 
\end{equation}
The intrinsic Helmholtz free energy functional, $F\left[ \rho\left(\vec{r}\right) \right]$, incorporates the properties of fluid. In general case, the functional depends on the external potential $V_{ext}\left(\vec{r}\right)$ if the intermolecular forces and/or intramolecular degrees of freedom are affected by the external field. However, we defer this discussion to the second part of the study, and consider $F\left[ \rho\left(\vec{r}\right) \right]$ invariant with respect to $V_{ext}\left(\vec{r}\right)$. The external potential describes the solid surfaces in contact with fluid, and induces inhomogeneiety in the system. The chemical potential $\mu$ is related to the ambient humidity.

The Helmholtz free energy functional $F\left[ \rho\left(\vec{r}\right) \right]$ can be presented as 
\begin{equation}
\label{F_contribs}
	F\left[ \rho\left(\vec{r}\right) \right] = F_{id}\left[ \rho\left(\vec{r}\right) \right] + \Delta F_{rep}\left[ \rho\left(\vec{r}\right) \right] + \Delta F_{attr}\left[ \rho\left(\vec{r}\right) \right],
\end{equation}
where $F_{id}\left[ \rho\left(\vec{r}\right) \right]$ corresponds to the ideal system without intermolecular interactions, and the other two terms describe additional intermolecular repulsion and attraction, respectively. The ideal part of the Helmholtz free energy functional is given explicitly by   
\begin{equation}
	F_{id}\left[ \rho\left(\vec{r}\right) \right] = kT\int \rho\left(\vec{r}\right) \left( \ln\left(\lambda^{3}_{th}\rho\left(\vec{r}\right)\right) -1 + \frac{\mu_{int}}{kT} \right) d\vec{r},
\end{equation}
$\lambda_{th}$ is a thermal wavelength, $k$ is the Boltzmann constant, $T$ is the temperature and the term $\mu_{int}\left(kT\right)^{-1}$ has been added to incorporate internal degrees of freedom of the water molecules.

The repulsive free energy functional is expressed as
\begin{equation}
	\Delta F_{rep}\left[ \rho\left(\vec{r}\right) \right] = kT\int \rho\left(\vec{r}\right) \Delta\psi_{hs}\left(\bar{\rho}\left(\vec{r}\right)\right) d\vec{r}.
\end{equation}
Here $\Delta\psi_{hs}$ is an excess free energy density for the repulsive fluid evaluated at every spatial point as a function of the smoothed (coarse-grained) density $\left(\bar{\rho}\left(\vec{r}\right)\right)$. The latter is an average of the local density  $\rho\left(\vec{r}\right)$, taken over a small domain using the weighting function $\omega$:
\begin{equation}
	\bar{\rho}\left(\vec{r}\right) = \int \rho\left(\vec{r'}\right) \omega\left( \left|\vec{r}-\vec{r'}\right|,\ldots \right) d\vec{r'}.
\end{equation}
The weighting function may depend not only on the spatial coordinate but also be a functional of $\rho$, depending on the particular non-local DFT formulation\cite{DFT_inhomog_fluids_Evans}. We use the repulsive intermolecular potential with the hard sphere diameter $\sigma$ and the Carnahan-Starling free energy density\cite{Carnahan_Starling_hsfluid} 
\begin{equation}
	\Delta\psi_{hs} = kT\frac{4\eta-3\eta^2}{\left(1-\eta\right)^2},
\end{equation}
where $\eta$ is the packing fraction of the hard spheres corresponding to the number density $\rho$: 
\begin{equation}
	\eta = \frac{\pi}{6} \sigma^3 \rho\left(\vec{r}\right). 
\end{equation}
For the sake of simplicity we choose a generalized van der Waals model\cite{DFT_inhomog_fluids_Evans, fluid_functionals_compar} with 
\begin{equation}
	\omega\left(\left|\vec{r}-\vec{r'}\right|\right) = \frac{3}{4\pi \sigma^3} \theta\left(\sigma-\left|\vec{r}-\vec{r'}\right|\right),
\end{equation}
where $\theta$ is a Heaviside step function. While the more sophisticated approaches, such as the Tarazona's model\cite{Tarazona_hs_functional}, have better accuracy for describing density profiles of the hard sphere systems, the choice of the generalized van der Waals model is justified in the systems with strong intermolecular attraction\cite{fluid_functionals_compar}, which is the case of our interest. 

In the framework of the mean field approximation, the third term in Eq.~(\ref{F_contribs}) describing an attractive part of the free energy functional can be written as:  
\begin{equation}
\label{Fattr}
	\Delta F_{attr}\left[ \rho\left(\vec{r}\right) \right] = \frac{1}{2}\int\int \rho\left(\vec{r}\right) \rho\left(\vec{r'}\right) \Phi_{attr}\left(\left|\vec{r}-\vec{r'}\right|\right) d\vec{r}d\vec{r'},
\end{equation}
where $\Phi_{attr}\left(\vec{r}\right)$ is the attractive pair intermolecular potential. We consider the water-like fluid for which $\Phi_{attr}\left(\vec{r}\right)$ is a square well potential characterized by two {\itshape effective} parameters such as the depth ($\epsilon$), and the width ($d$) of the well. This choice is related to the fact that in water, the short-range hydrogen bonding dominates the longer-range dipole-dipole and higher order interactions. Similar forms of attractive potentials were used before\cite{hbond_fluid_Wertheim, water_model_BenNaim}.  

The minimization of the functional given by Eq.~(\ref{GCOmega}) with $F\left[ \rho\left(\vec{r}\right) \right]$ presented above by Eqs.~(\ref{F_contribs})-(\ref{Fattr}) leads to the following integral equation for the equilibrium density profile $\rho\left(\vec{r}\right)$:
\begin{widetext}
\begin{subequations}
\label{int_eqn_for_rho}
\begin{equation}
	kT\left(\ln\left(\lambda^{3}_{th}\rho\left(\vec{r}\right)\right)-1\right) + \mu_{int} + \Delta\psi_{hs}\bar{\rho}\left(\vec{r}\right) + J_{hs}\left[\rho\left(\vec{r}\right),\bar{\rho}\left(\vec{r}\right)\right] + J_{int}\left[\rho\left(\vec{r}\right)\right] + V_{ext}\left(\vec{r}\right) = \mu,
\label{int_eqn_intelf}
\end{equation}
\begin{equation}
	J_{hs}\left[\rho\left(\vec{r}\right),\bar{\rho}\left(\vec{r}\right)\right] = \frac{3}{4\pi\sigma^3}\int \Delta\psi'_{hs}\left(\bar{\rho}\left(\vec{r'}\right)\right) \theta\left(\sigma-\left|\vec{r}-\vec{r'}\right|\right) \rho\left(\vec{r'}\right) d\vec{r'},
\label{Jhs}
\end{equation}
\begin{equation}
	J_{int}\left[\rho\left(\vec{r}\right)\right] = -\epsilon\int \theta\left(\left|\vec{r}-\vec{r'}\right|-\sigma\right) \theta\left(d-\left|\vec{r}-\vec{r'}\right|\right) \rho\left(\vec{r'}\right) d\vec{r'},
\label{Jint}
\end{equation}
\end{subequations}
\end{widetext}
where a prime at $\Delta\psi_{hs}$ in (\ref{Jhs}) denotes the differentiation with respect to $\rho$. The Eq.~(\ref{int_eqn_intelf}) is to be solved for $0\leq r\leq L$, $0\leq z\leq t+R-\sqrt{R^2-r^2}$, and $0\leq\phi\leq2\pi$. The boundary $r = L$ is chosen such that the region within which the liquid bridge may be formed corresponds to $r\ll L$. 
 
An explicit form of the external potential, $V_{ext}\left(\vec{r}\right)$ describing interaction of the fluid with the confining surfaces is the final piece of the information required for closure of the model's description. For that we use a two-parametric exponential attraction potential on top of the hard wall repulsion:
\begin{equation}
	V_{ext}\left(x\right) = \left\{
\begin{array}{cc}
-\epsilon_s \exp\left(-\alpha x\right), & x\geq0\\
\infty, & x<0
\end{array} \right\}
\label{Vext}
\end{equation}
where $x$ is the distance from the surface. The parameter $\epsilon_s$ is the work required to displace a water molecule from $x=0$ to infinity corresponding to the binding energy of water at given surface.

\subsection{Computational part}

The triple integrals (\ref{Jhs}) and (\ref{Jint}) must be evaluated over 3-D volume. Fortunately, the character of their angular dependence in a step-function form allows reducing them to the double integrals below:
\begin{subequations}
\begin{equation}
	\int\int\int u\left(\vec{r}\right) \theta\left(p-\left|\vec{r}-\vec{r_0}\right|\right) d\vec{r} = \int rdr \int dz\cdot2\lambda u\left(r,z\right),
\end{equation}
\begin{equation}
	\lambda = \left\{
\begin{array}{cl}
\arccos\left(\Lambda\right), & -1<\Lambda<1\\
0, & \Lambda\geq 1\\
\pi, & \Lambda\leq -1
\end{array} \right\},
\end{equation}
\begin{equation}
	\Lambda = \frac{\left(z-z_0\right)^2+r^2+r^2_0-p^2}{2rr_0}.
\end{equation}
\end{subequations}

The iterative method used to solve nonlinear integral Eq.~(\ref{int_eqn_for_rho}) deserves special description. Simple successive approximation procedure normally fails for the equations similar to (\ref{int_eqn_for_rho}) even for one-dimensional problems as the computation diverges after the few iterations\cite{Tarazona_hs_functional, fluid_functionals_compar}. Several approaches were used to overcome this difficulty for the one-dimensional case. Tarazona\cite{Tarazona_hs_functional} used the mixing of the new and previous iterative values with a space-dependent mixing function. While this approach fixes divergence, it requires a careful selection of the mixing function through trial end error, and converges slowly. Vanderlick {\itshape et.al.}\cite{fluid_functionals_compar} used a uniform discretization of the domain and converted an integral equation into the system of nonlinear algebraic equations. This works well for one-dimensional problems, but is subjective to the choice of the discretization method. 

In our model, the spatial domain for which Eq.~(\ref{int_eqn_for_rho}) has to be solved is two dimensional and asymmetric, which requires a general approach. Below we suggest an iterative scheme conceptually similar to Newton's iterative method for the systems of algebraic equations. The general form of Eq.~(\ref{int_eqn_for_rho}) can be written as:
\begin{equation}
	\ln\rho + \Phi\left(\rho\right) = B,
\end{equation}
where $\Phi\left(\rho\right)$ has a complicated dependence with respect to $\rho$. The iterative scheme is set up in the form
\begin{equation}
	\ln\rho^{(m+1)} + \Phi\left(\rho^{(m)}\right) + \left(\rho^{(m+1)}-\rho^{(m)}\right)\left(\frac{\partial\Phi}{\partial\rho}\right)_{\rho^{(m)}} = B.
\label{iter_sch}
\end{equation}
Here $\rho^{(k)}$ is an approximation for $\rho\left(r,z\right)$ obtained at $k^{th}$ iteration. The integral terms appearing in $\Phi$ and its derivative are calculated simultaneously. The equation to be solved at $m^{th}$ iteration has the following form:
\begin{equation}
	\ln\rho^{(m+1)} + D\rho^{(m+1)} = C,
\end{equation}
and its solution can be presented via the Lambert W-function:
\begin{equation}
	\rho^{(m+1)} = \exp\left(C-W\left(De^C\right)\right).
\label{iter_W}
\end{equation}
We found that the iterative procedure given above through Eqs.~(\ref{iter_sch})-(\ref{iter_W}) is very efficient so that the convergence is often accomplished in less than 10-20 iterations.

\section{Results and Discussion}
\label{sec:results}

\subsection{Bulk fluid and model parameters}
\label{subsec:bulk}

For the case of the uniform bulk fluid, Eq.~(\ref{int_eqn_for_rho}) reduces to the following form: 
\begin{eqnarray}
		\mu = kT\left(\ln\left(\lambda^{3}_{th}\rho\right)-1\right) + \mu_{int} + \Delta\psi_{hs} + \rho\Delta\psi'_{hs} \nonumber\\ - \frac{4\pi}{3}\left(\left(\frac{d}{\sigma}\right)^3-1\right)\rho\sigma^3\epsilon.
\label{mu_bulk}
\end{eqnarray}
To determine the parameters $\sigma$, $d$ and $\epsilon$ in Eq.~(\ref{mu_bulk}) we use the characteristics of water at the critical point determined by the following thermodynamic conditions\cite{book_phase_trans_Stanley}:
\begin{equation}
	\left(\frac{\partial\mu}{\partial\rho}\right)_{T_c} = 0, \;\;\; \left(\frac{\partial^2\mu}{\partial\rho^2}\right)_{T_c} = 0. 
\end{equation}
The derivatives are evaluated at the critical temperature $T_c$ (647 K for water). The solution of these equations with $\mu$ explicitly given by Eq.~(\ref{mu_bulk}) determines two groups of the parameters:
\begin{equation}
	\sigma^3\rho_c = 0.2372, \;\;\; \frac{4\pi}{3}\left(\left(\frac{d}{\sigma}\right)^3-1\right) \frac{\epsilon}{kT_c} = 10.3882.
\end{equation}
The density of water under critical point conditions, $\rho_c$, results in $\sigma=2.8$\AA, which is commonly taken as a hard sphere diameter of water molecule\cite{water_model_BenNaim}. Separation of the parameters $\epsilon$ and $d$ in the second group is not necessary as they are always grouped together in this form.

The second consideration for the bulk fluid deals with liquid-vapor coexistence at the temperature $T = 300\;K$. Two coexistence equations are derived assuming the chemical potential and pressure equal for both phases. The chemical potential is given by Eq.~(\ref{mu_bulk}), and the pressure is presented as a function of number density, $P\left(\rho\right)$, using the expression for the grand canonical potential, $\Omega$, for the bulk fluid: 
\begin{eqnarray}
		-\frac{\sigma^3}{kT}\;P = \sigma^3\rho\left( Q + \ln\left(\sigma^3\rho\right) + \frac{\Delta\psi_{hs}}{kT} \right) \nonumber\\ - \frac{2\pi}{3} \left(\left(\frac{d}{\sigma}\right)^3-1\right) \frac{\epsilon}{kT} \left(\sigma^3\rho\right)^2,
\end{eqnarray}
where dimensionless parameter $Q$ was introduced as
\begin{equation}
	Q = 3\ln\left(\frac{\lambda_{th}}{\sigma}\right)-1+\frac{\mu_{int}}{kT}. 
\end{equation}
After solving the equations above, we find $\sigma^3\rho_v=1.86\times10^{-4}$ and $Q = 5.6$ selecting the density of water in the liquid phase at 300 K, $\sigma^3\rho_l=0.731$. The vapor phase density $\rho_v$ is used as a reference value to calculate chemical potential $\mu$ for a given relative humidity $H_r$.

\subsection{Liquid meniscus condensation in the nanoscale junction}
\label{subsec:menisc}

\begin{figure}
\includegraphics[width=8.5cm]{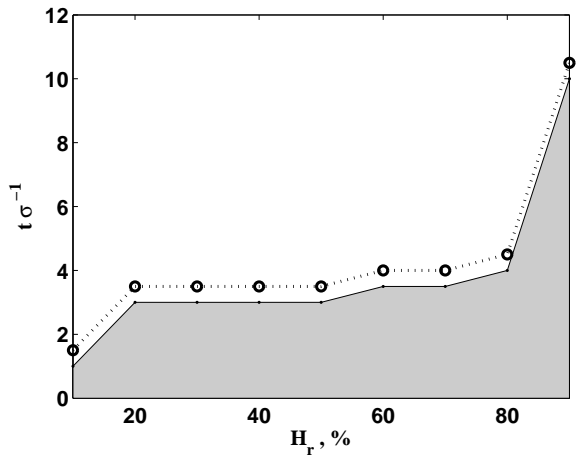}
\caption{Water condensation between planar surface and spherical asperity of the radius $R=36\sigma$ (10 nm). Shaded area corresponds to the formation of liquid meniscus. No liquid bridge is formed for given $H_r$ when $t$ reaches ordinate of the open circle. The parameters of fluid interaction with confining walls are $\epsilon_s\epsilon^{-1}=1$ and $\alpha\sigma=2$.}
\label{fig:liq_region}
\end{figure}

The fluid density distribution was calculated, using the methodology above, in the geometry illustrated in Fig.~\ref{fig:geom}, for different values of relative humidity $H_r$ and separation between confining walls $t$, at $T = 300\;K$. It was found that for certain values of $H_r$ and $t$, the fluid tends to condense in the region near $r=0$, forming a liquid meniscus that connects the confining surfaces. The tendency for liquid meniscus condensation increases with $H_r$ and diminishes with the growth of $t$. The region in $H_r-t$ coordinates, corresponding to the meniscus formation, is depicted in Fig.~\ref{fig:liq_region}.    

\begin{figure}
	\begin{minipage}{8cm}
		\begin{center}
		\includegraphics[width=7.5cm]{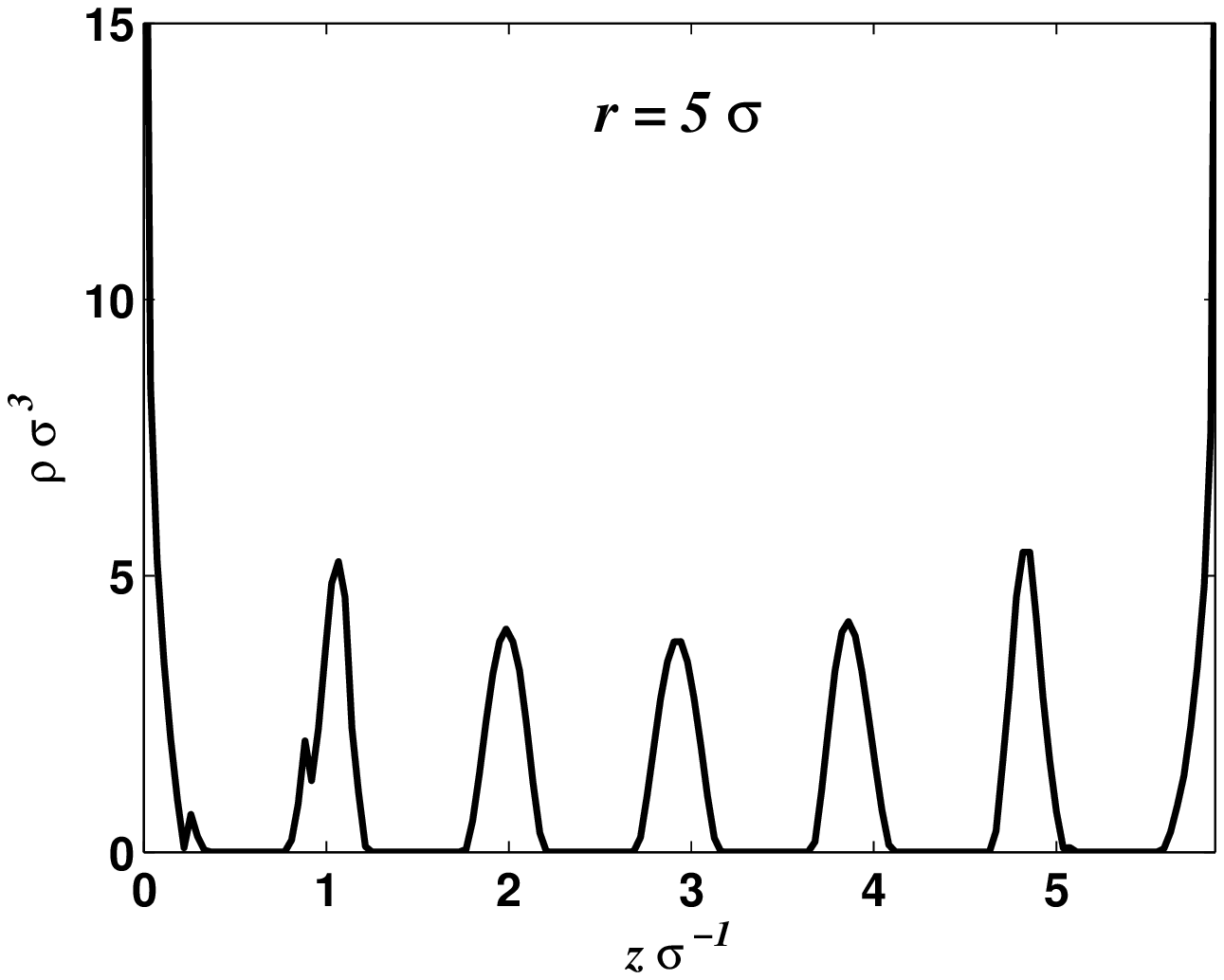}
		\end{center}
	\end{minipage} 
	\begin{minipage}{8cm}
		\begin{center}
		\includegraphics[width=7.5cm]{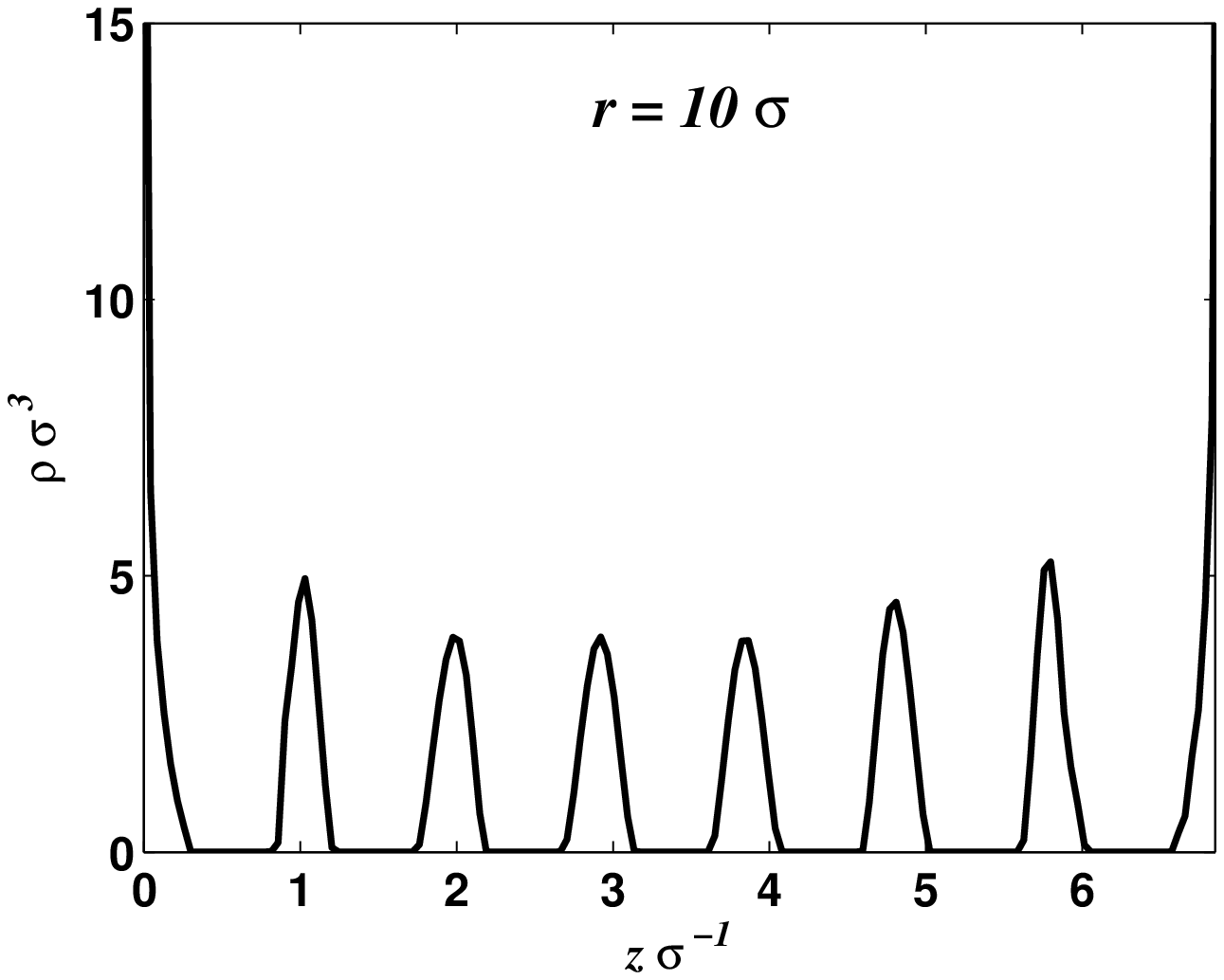}
		\end{center}
	\end{minipage} 
	\begin{minipage}{8cm}
		\begin{center}
		\includegraphics[width=7.5cm]{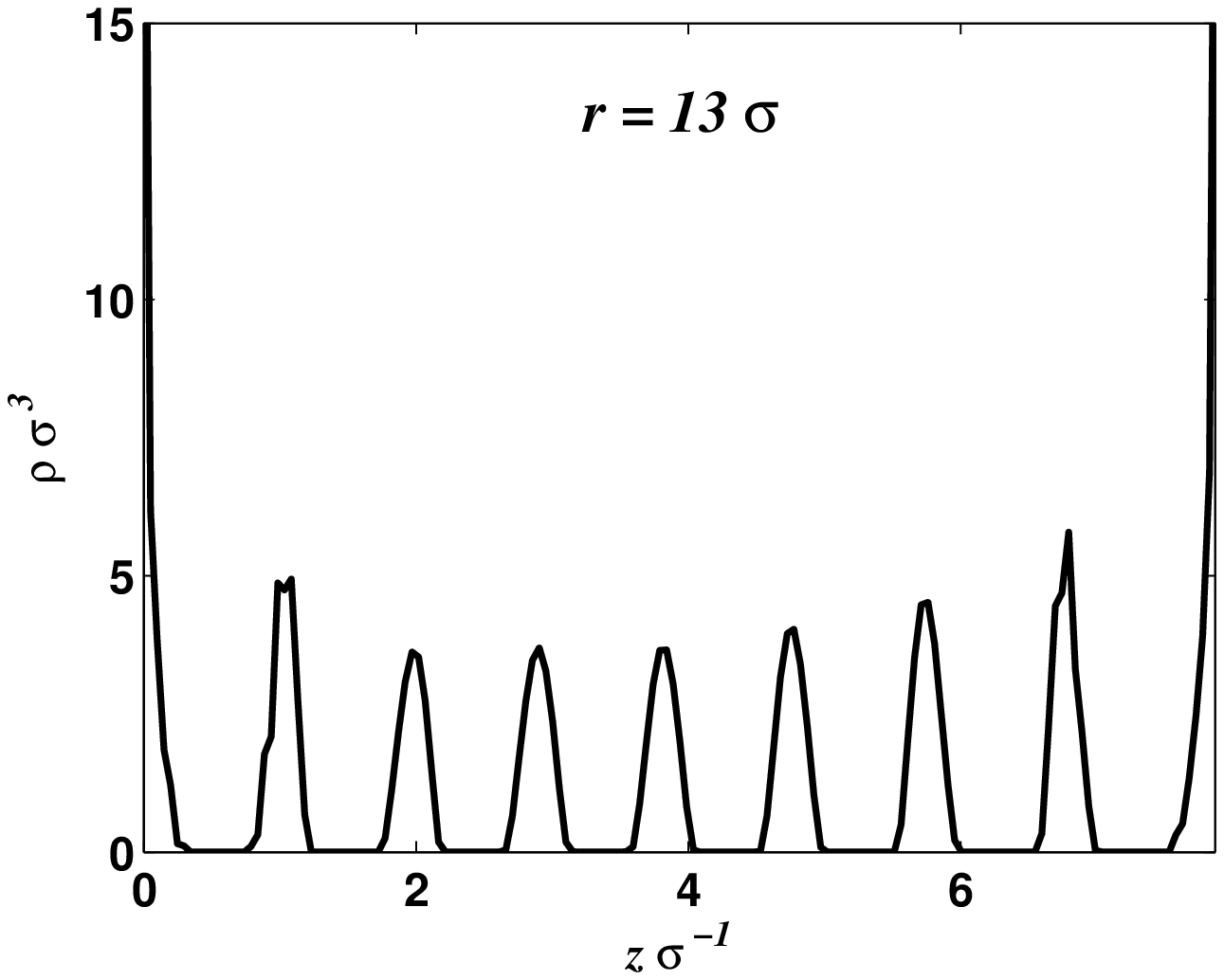}
		\end{center}
	\end{minipage} 
\caption{Fluid density profiles between planar surface and spherical asperity of the radius $R=36\sigma$ (10 nm), separated by $t=5.5\sigma$, for three different values of the radial coordinate $r$. Relative humidity is 90\%, and parameters of fluid interaction with solid walls are $\epsilon_s\epsilon^{-1}=1$ and $\alpha\sigma=2$.}
\label{fig:rho_prof}
\end{figure}

Fig.~\ref{fig:rho_prof} presents typical density profiles: $\sigma^3\rho$ as a function of the axial coordinate $z\sigma^{-1}$, for different radial positions $r$. The density profile for a given $r$ is mainly determined by the separation between the walls at that point, providing that the fluid-wall interaction parameters are fixed. Inside the meniscus, the density is higher near the walls and exhibits maxima in-between (Fig.~\ref{fig:rho_prof}). Significant density redistribution occurs near the meniscus boundary. For the longer radial distances, the liquid wets confining surfaces according to the specified values of $\epsilon_s$ and $\alpha$.

An essential question is how to define the size of liquid meniscus based on the computed fluid density distribution, or what factors define liquid/vapor boundary for a smooth density distribution in the asymmetric geometry. The answer can be found should we specify the phenomena of interest and its description based on the dimensions of the liquid bridge. The description of the charge transport in nanoscale junction and calculation of the forces exerted by fluid on the confining walls may require the usage of different effective meniscus radii. A certain approximation is required since this work studies general peculiarities of meniscus formation under variable conditions. In earlier work related to meniscus formation\cite{menisc_MC_Ratner_JCP02}, an {\itshape ad hoc} chosen threshold of half filled lattice site has been used. In this work, we use a single threshold value of the number density to separate liquid and vapor regions, and define the boundary. We assume that if the interaction energy of water molecule located at a given point of space with its neighbors is equal or less than $kT$, then it belongs to vapor phase. The corresponding quantitative criterion can be presented using Eq.~(\ref{Jint}) as
\begin{equation}
	-J_{int}\left(\rho\right) = kT,
\end{equation}
or
\begin{equation}
	\sigma^3\rho = \frac{3kT}{4\pi\epsilon} \left( \left(\frac{d}{\sigma}\right)^3 -1 \right)^{-1}.
\label{rho_th}
\end{equation}
This results in the threshold value between the liquid and vapor phases to be $\sigma^3\rho_{th} = 0.16$. For comparison, we found that the bulk liquid density was $\sigma^3\rho_l = 0.731$, while the vapor density was $\sigma^3\rho_v = 1.86\times10^{-4}$ (as described in Section~\ref{subsec:bulk} above).

\begin{figure}
\includegraphics[width=8.5cm]{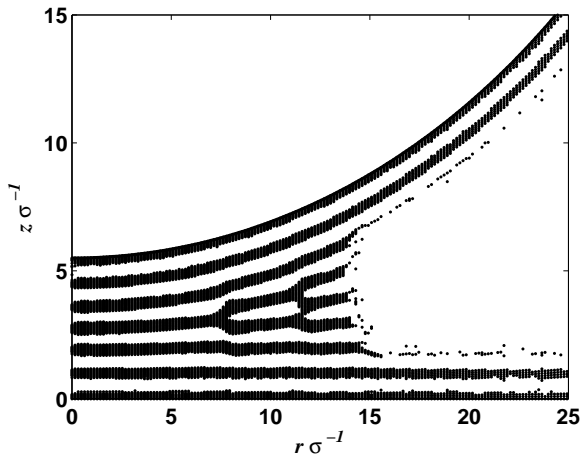}
\caption{Water meniscus formed between planar surface and spherical asperity of the radius $R=36\sigma$ (10 nm), separated by $t=5.5\sigma$. Relative humidity is 90\%, and parameters of fluid interaction with solid walls are $\epsilon_s\epsilon^{-1}=1$ and $\alpha\sigma=2$.}
\label{fig:menisc_prof}
\end{figure}

Fig.~\ref{fig:menisc_prof} presents the liquid meniscus that corresponds to the density profiles plotted in Fig.~\ref{fig:rho_prof}. The spaces on the plot filled with dots correspond to the regions (domain discretization points) for which the fluid density exceeds the threshold value $\rho_{th}$ defined above.

\begin{figure}
	\begin{minipage}{7cm}
		\begin{center}
		\includegraphics[width=7cm]{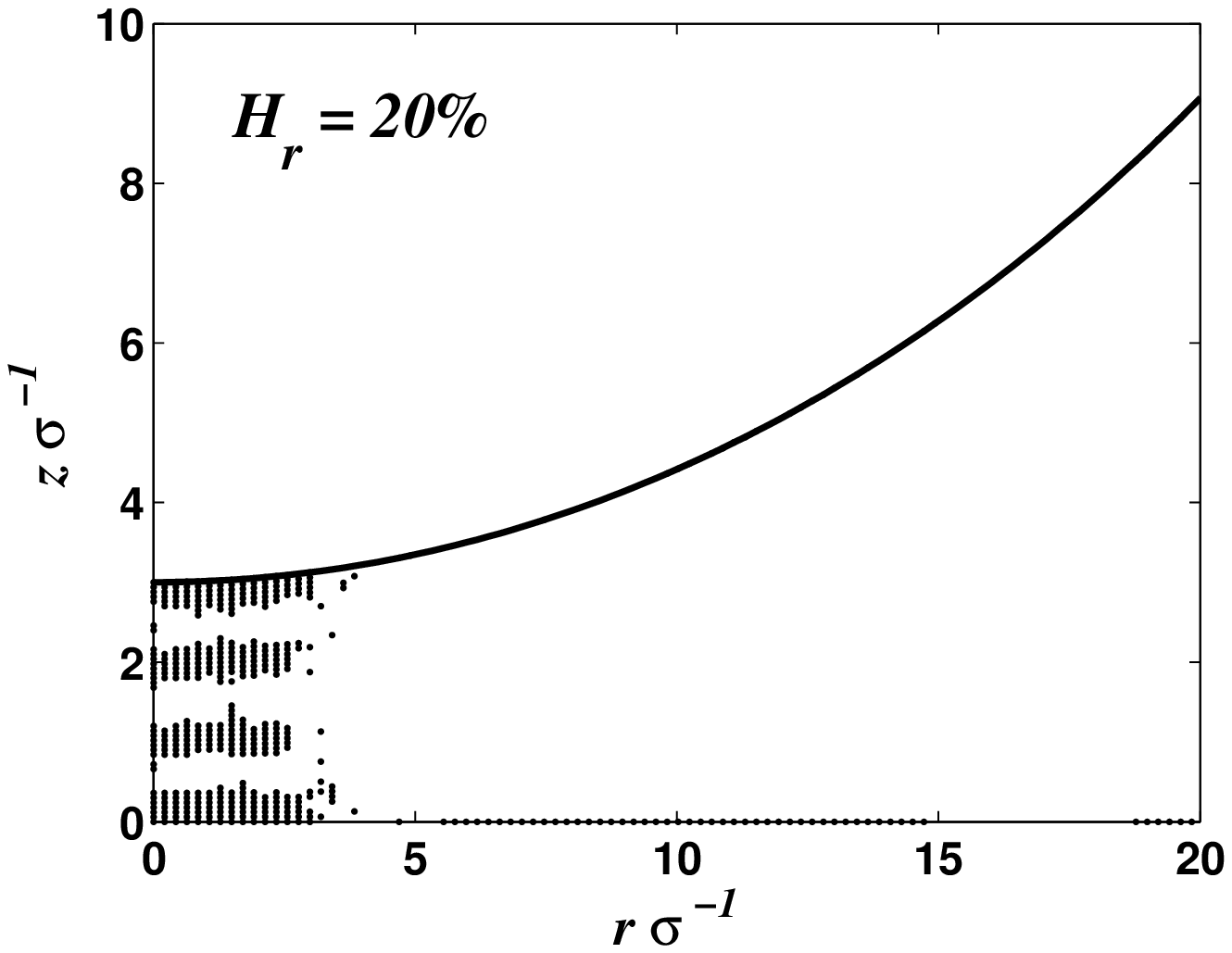}
		\end{center}
	\end{minipage} 
	\begin{minipage}{7cm}
		\begin{center}
		\includegraphics[width=7cm]{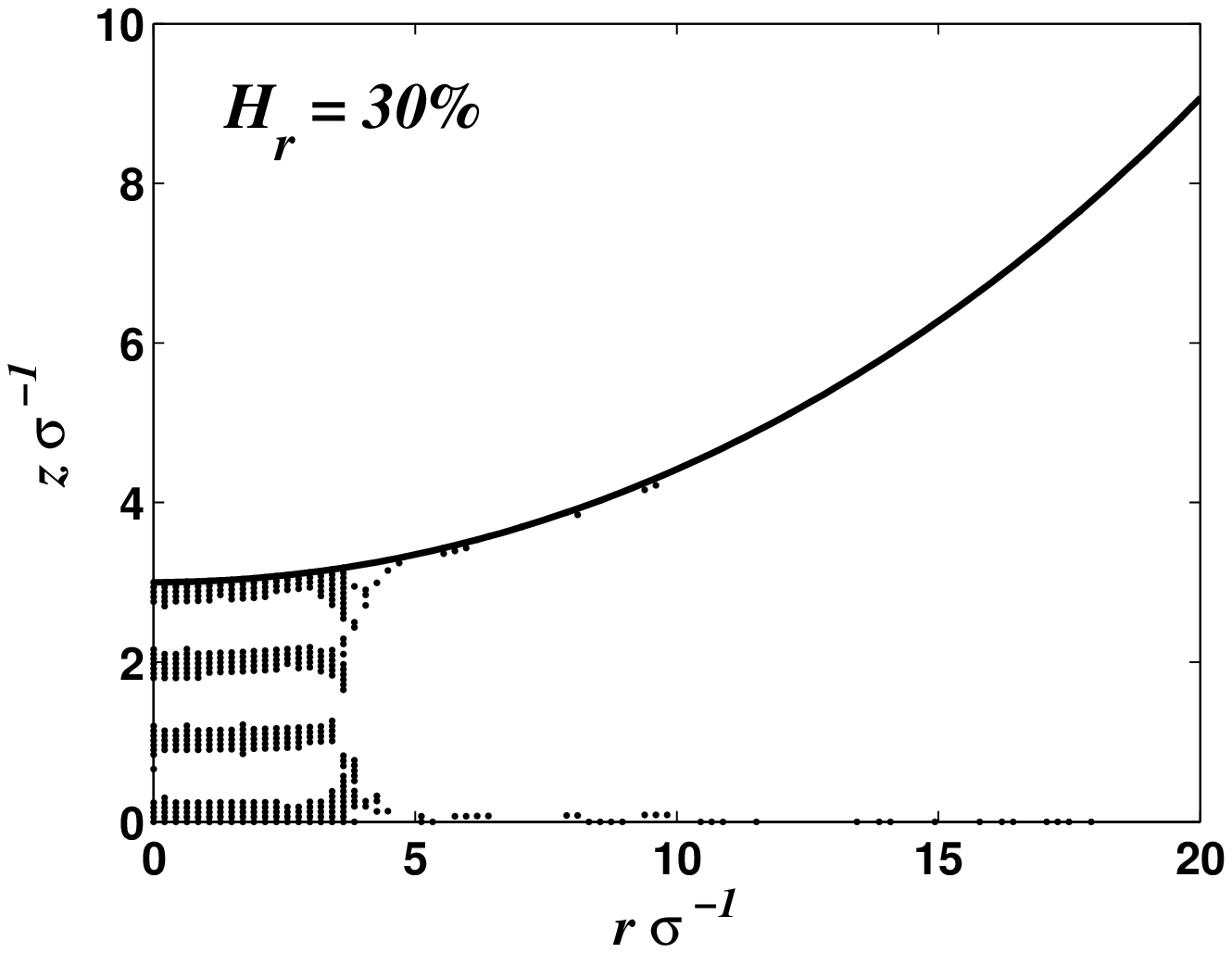}
		\end{center}
	\end{minipage} 
	\begin{minipage}{7cm}
		\begin{center}
		\includegraphics[width=7cm]{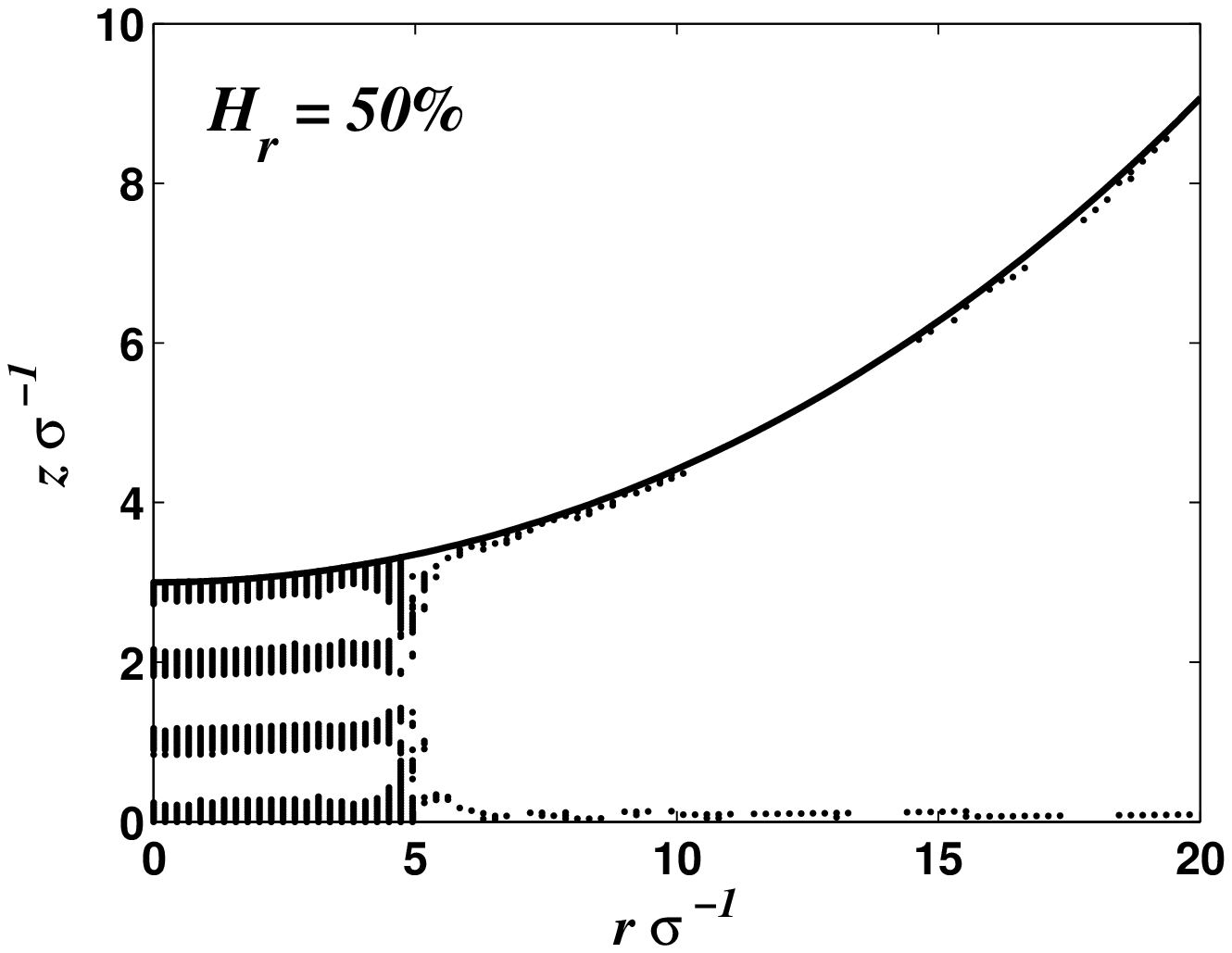}
		\end{center}
	\end{minipage} 
\caption{Variation of the meniscus size with relative humidity $H_r$. Filled space on the plots corresponds to the regions where fluid density exceeds the threshold value of $\sigma^3\rho_{th}=0.16$. Other relevant parameters are: $t=3\sigma$, $\epsilon_s\epsilon^{-1}=1$ and $\alpha\sigma=2$. Curvature radius of the asperity is $R=36\sigma$.}
\label{fig:menisc_Hr}
\end{figure}

Fig.~\ref{fig:menisc_Hr} illustrates the variation of the meniscus's size with respect to the relative humidity $H_r$. Again, the spaces on the plots filled with dots correspond to the regions for which the fluid density exceeds $\rho_{th}$. The meniscus boundary shifts to larger values of $r$ as humidity grows.

\begin{figure}
	\begin{minipage}{7cm}
		\begin{center}
		\includegraphics[width=7cm]{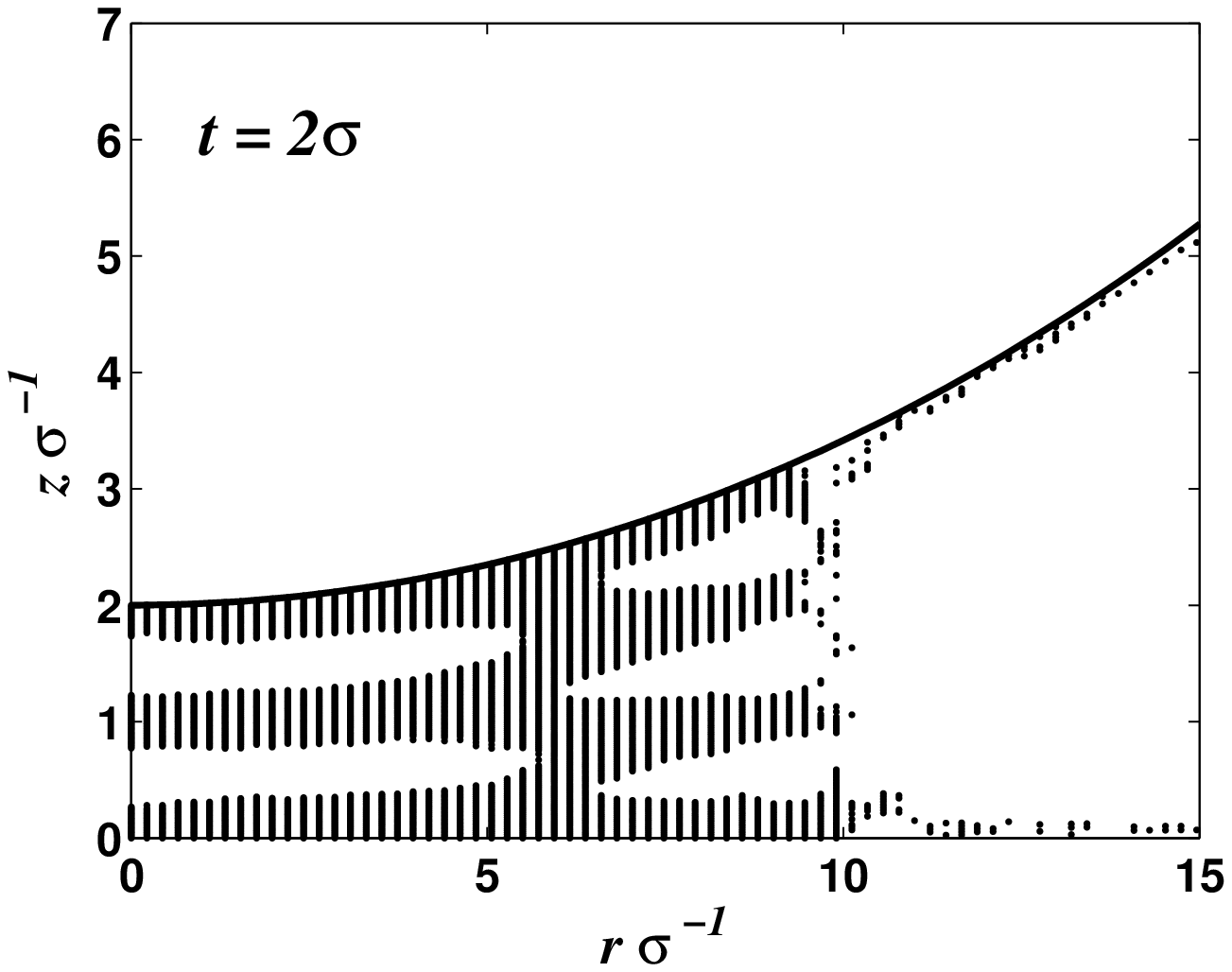}
		\end{center}
	\end{minipage} 
	\begin{minipage}{7cm}
		\begin{center}
		\includegraphics[width=7cm]{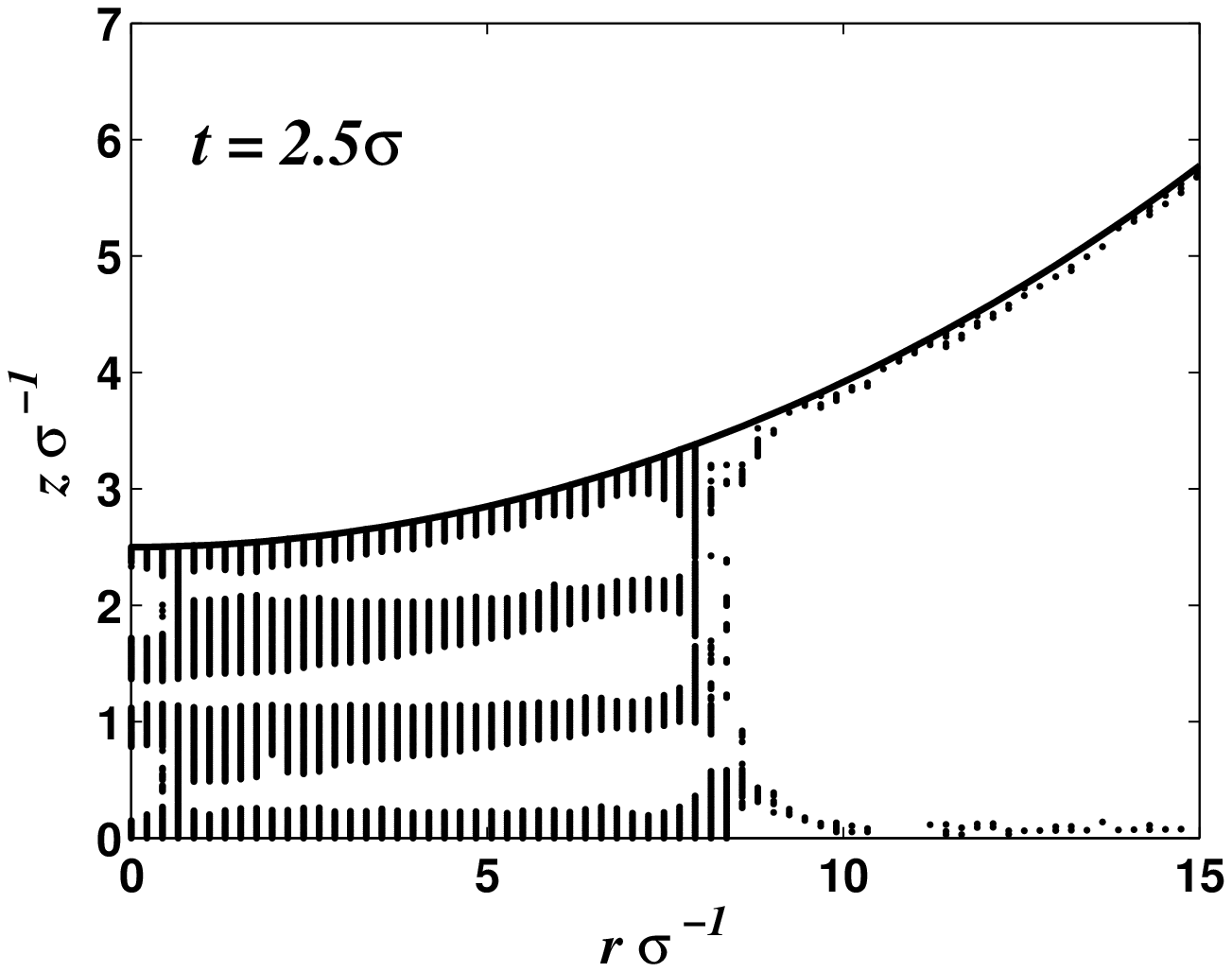}
		\end{center}
	\end{minipage} 
	\begin{minipage}{7cm}
		\begin{center}
		\includegraphics[width=7cm]{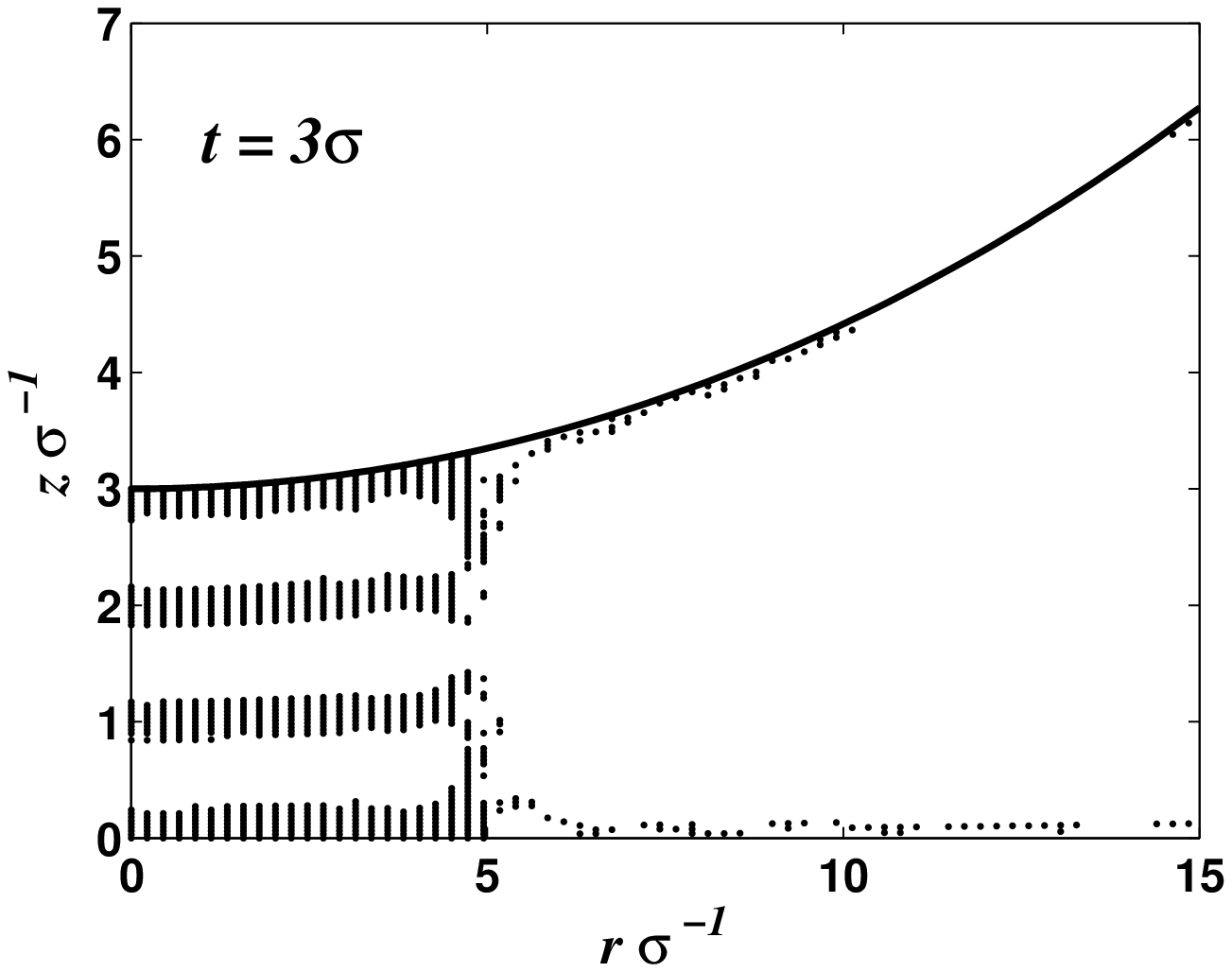}
		\end{center}
	\end{minipage} 
	\begin{minipage}{7cm}
		\begin{center}
		\includegraphics[width=7cm]{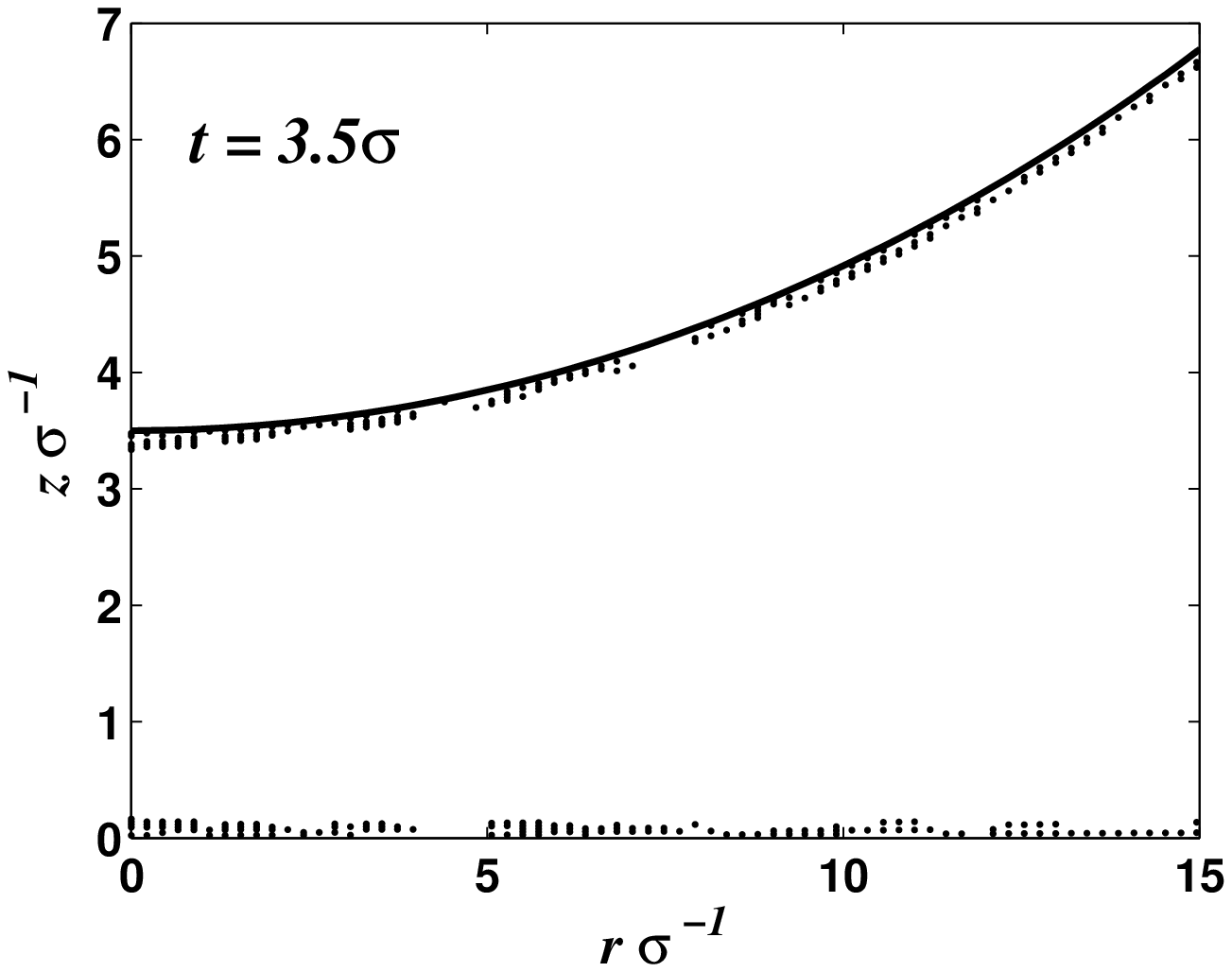}
		\end{center}
	\end{minipage} 
\caption{Variation of the meniscus size with separation $t$ between confining surfaces. Filled space on the plots corresponds to the regions where fluid density exceeds the threshold value of $\sigma^3\rho_{th}=0.16$. Relative humidity is 50\%, and parameters of fluid interaction with solid walls are $\epsilon_s\epsilon^{-1}=1$ and $\alpha\sigma=2$. Curvature radius of the asperity is $R=36\sigma$.}
\label{fig:menisc_t}
\end{figure}

Fig.~\ref{fig:menisc_t} shows similar plots for the variable separation $t$ between the walls. The meniscus narrows (as can be seen for $t\sigma^{-1}=2-3$) and eventually disappears ($t\sigma^{-1}=3.5$) as $t$ increases.

\subsection{Fluid - wall interaction parameters: An Estimation}
\label{subsec:fluid_wall_inter}

The parameters, $\epsilon_s$ and $\alpha$ (Eq.~(\ref{Vext})) characterizing the energy of the fluid interaction with solid surfaces ($\epsilon_s$), and the spatial range ($\alpha$) of the corresponding potential, must be estimated for specific surfaces to apply the results of the calculations to real physical systems. We suggest using the polymolecular adsorption isotherms for this purpose since the data of water adsorption is more abundant than those related to other measurements.

The fluid-wall interaction potential (\ref{Vext}) is two-parametric. First, we restrict the potential to act in the region of thickness about $\sigma$ near the surface by setting $\alpha\sigma=2$. Then the adsorption isotherms for the model fluid on surfaces with different values of $\epsilon_s$ are calculated.

\begin{figure}
\includegraphics[width=8cm]{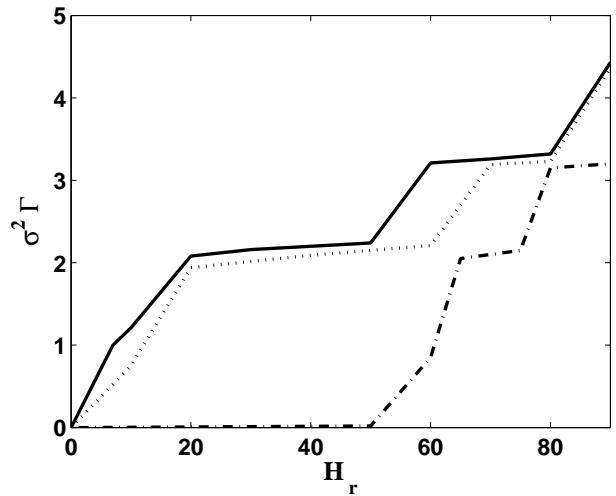}
\caption{Calculated adsorption isotherms for the model fluid on surfaces with different values of $\epsilon_s\epsilon^{-1}:$ 1 (dash-dot line), 1.5 (dotted line), 2 (solid line), and with $\alpha\sigma=2$. $\Gamma$ is number of molecules adsorbed per unit surface area. As $H_r$ approaches 100\%, the onset of (integrable) singularity is observed (not shown).}
\label{fig:adsorpt}
\end{figure}

The number of molecules adsorbed per surface area of $\sigma^2$ (integrated or Gibbs adsorption) with respect to the relative humidity $H_r$ for $T = 300 K$ is presented in Fig.~\ref{fig:adsorpt}. A singularity is observed (not shown in the figure) as the humidity ($H_r$) approaches 100\%, consistent with thermodynamical requirement\cite{adsorpt_thermodyn_restr}. The isotherms exhibit typical step-like structure ($\epsilon_s\epsilon^{-1}\sim 1$) for the intermediate humidity values, similar to the results of calculations based on three-dimensional Ono-Kondo lattice model\cite{polymol_ads_interact_simul} and experimental data for multilayer adsorption of water\cite{ads_water_NaCl_surfsci}. The parameter $\epsilon_s$ for specific surface of interest may thus be evaluated by fitting the adsorption isotherm.  
 
The above considerations imply both geometric and energetic uniformity of the surface. The geometric one means the average surface roughness is much smaller than the separation distance ($t$) between surfaces. The latter assumption can be adapted to energetically nonuniform surfaces by taking the effective (average) value of $\epsilon_s$ and requiring that the radial region of interest is large enough to sample most of the adsorption energy distribution. Additionally, it is assumed that no significant dissociation of water occurs on the surface, and adsorption is reversible.  

\section{Summary}
\label{sec:summary}

A spatial distribution of the water-like fluid in the asymmetric nanoscale junction formed in proximity of a nanoscale asperity has been studied using an approach based on non-local density functional theory. The approach presents a closed set of procedures modeling the fluid resembling water. The computational part of the approach is adapted for asymmetric geometry of the model comprising two solid surfaces. The basic framework for the density functional description of the case with electric field induced anisotropy in the fluid has been developed. The route for quantitative predictions on the water meniscus formation near atomic force microscope tip dwelling above a sample surface is one of the important applications of the developed methodology.  

Our results suggest further development of the modeling in three directions. The first direction would be a self-consistent description for electrically charged or biased asperity comprising electric field and fluid density distributions. The second direction would be the calculation of forces acting on the nano-asperity based on the density functional formalism, which would have applications in the context of AFM-assisted nanolithography. The third direction would concentrate on the study of dynamical phenomena at the nano-asperity including the kinetics of capillary condensation under highly nonuniform external potentials. 

\begin{acknowledgments}
This work was supported through Air Force Office Sponsored Research grant F49620-02-1-428 in the frames of Akron/Air Force Center of Polymer Photonics.
\end{acknowledgments}

\bibliography{bio}

\end{document}